# First-principles calculations and experimental studies on $Co_2FeGe$ Heusler alloy nanoparticles for spintronics applications

Aquil Ahmad[1], S. K. Srivastava[1], and A. K. Das[1]

[1]*Department of Physics, Indian Institute of Technology Kharagpur, Kharagpur, India-721302*

*Email(s): aquil@phy.iitkgp.ernet.in (A. Ahmad); sanjeev@phy.iitkgp.ernet.in (S.K. Srivastava); amal@phy.iitkgp.ernet.in (A.K. Das)*

## ABSTRACT

Here, we report the synthesis and physical properties of $Co_2FeGe$ (CFG) Heusler alloy (HA) nanoparticles (NPs). The NPs of size 23 ± 10 nm are prepared using the co-precipitation method. X-ray and selected area electron diffraction patterns have confirmed the cubic Heusler phase of the NPs with the A2-disorder. These NPs are soft ferromagnetic, and exhibit a high saturation magnetization ($M_s$) along with a very high Curie temperature ($T_c$) of 1060 K. The observed $T_c$ value matches closely with the theoretically calculated one following a model provided by Wurmehl *et al.* [1]. The high $M_s$ and $T_c$ make the present system a potential candidate for magnetically activated nano-devices working at high temperatures. The near-integral value 5.9 $\mu_B$/f.u. of $M_s$ at low temperatures indicates that the half-metallic ferromagnetism is preserved even in the particles even on the 20 nm length scale. Additionally, we have facilitated the existing HA-NP preparation method, which can be used in synthesizing other HA-NPs. The first-principles density functional theory computations complement the experimental results.

## 1. Introduction

Magnetic nanoparticles (NPs) have drawn an extensive interest in different scientific fields [2]. They have demonstrated novel physical properties because of their high surface to volume ratio. Nanoscale magnetic materials are potential candidates for medical imaging, permanent magnets, and ultrahigh-density magnetic recording, most of them being binary alloys [2-6]. Extensive efforts have been made to synthesize and characterize magnetic NP's. However, such studies are mostly limited to core-shell nanomaterials [7], binary alloys [8], and oxides [9]. Recently, Heusler based intermetallic compounds [10] have received a considerable attention owing to their exceptional physical properties, such as high spin polarization (P) up to 100%, half-metallic ferromagnetism (HMF), high Curie temperature ($T_c$) and high magnetic moment [11-16]. The readers are referred to our previous paper in order to understand the detailed mechanism of half-metallic ferromagnetism in $X_2YZ$ type full-HAs [13]. The importance of such materials relies on the fact that their physical properties are easily tunable in an unusual way for the novel application in spintronics [10, 17], shape-memory effects [18, 19], thermoelectrics [20, 21], topological insulators [22-24] and magnetic refrigeration [25, 26]. Some recent studies also show that they can be used as potential catalysts [27, 28]. Heusler-based compounds have intensively been used in thin film form in numerous spin-based and microelectronics devices [29-32]. Heusler-based nanoparticles (NPs) and nanowires (NWs) are envisaged to be exploitable in innumerable applications, such as spintronics, topological insulators, and magnetic cooling [33-35]. In recent years, several attempts have been made to synthesize Heusler-based NPs and NWs. For example, C. Tsai *et al*. [36] have grown single crystal NWs of CoSiGe alloy via spontaneous chemical vapor transport method and investigated the effects of partial substitution of Ge for Si on the electrical transport, magnetic properties, and magnetoresistance. K. R. Sapkota *et al*. have synthesized $Co_2FeAl$ NWs having B2 or A2 disordered phase using electrospinning method. P. Simon *et al*. [37] have prepared $Co_2FeGa$ nanowires using silica template-aided method, showing the ideal $L2_1$ crystalline phase. S. Khan *et al.* [38, 39] prepared $Fe_2CoSn$ and $Co_2Mn_{0.5}Fe_{0.5}Sn$ NWs via electrodeposition technique and studied their structural, magnetic and transport properties. Lately, L. Galdun *et al*. [40] have reported the first electrochemical synthesis of $Co_2FeIn$ nanowires with a mean diameter 180 nm and length ~ 14.5 µm. These nanowires were fabricated by a simple template-assisted electrodeposition method, which can also be used in preparing other Heusler nanomaterials. L. Basit *et al*. [41] synthesized $Co_2FeGa$ nanoparticles having mean diameter ~20 nm by a chemical method. The magnetic moment of these $L2_1$ ordered NPs was ~ 5 $\mu_B$, which is the same as in the corresponding bulk. This suggests that the HMF remains unaltered even at this nanodimensional length scale. Lately, C. Wang *et al*. investigated the size dependent physical properties of CoNiGa nanoparticles [42], wherein an increment in the saturation magnetization ($M_s$) and $T_c$ was observed with increasing particle size. It is to be noted that the authors in [41, 42] have used fumed silica as a template in order to grow and control the size of the particles. On the other hand, Yang *et al*. [43] have reported a simple and effective chemical route – the so-called co-precipitation method – to prepare Heusler nanoalloys. Here, the authors studied the influence of the pH value and composition on physical properties of Co-Fe-Al Heusler nanoparticles. Nehla *et al*. [44] studied the structural, magnetic and transport properties of $Co_2FeGa$ nanoparticles. Recently, $Co_2/Fe_2$ based Heusler NPs have been synthesized by A. Ahmad *et al*., Duan *et al*., and Saravanan *et al*. [16, 45-48]. They all have studied the structural, magnetic, and magnetocaloric properties of the NPs. Some other Heusler nanoparticles have also been synthesized and studied recently [49-51].

The synthesis of Heusler alloy nanoparticles with desired stoichiometry and phase is challenging due primarily to lattice mismatches and immiscibilities. Therefore, previous studies have focused mostly on mere syntheses and characterizations. In HAs, structural disorder and deviation of magnetic moments from an integer value are shown to affect their half-metallicity [10, 13, 15]. Furthermore, their size reduction down to nanometric scale increases the structural disorder. Therefore, it is imperative to investigate how the system's structural disorder might affect its physical properties. A size-controlled nanometric sample could act as a perfect model. To our knowledge, this is the first report on the synthesis and study of structural and magnetic characteristics of $Co_2FeGe$ NPs. Besides, our modified methodology has facilitated the synthesis process, which can be suitable for synthesizing other Heusler based nanoalloys. The phase analysis of $Co_2FeGe$ nanoparticles (CFG-NPs) reveals that the present sample is crystallized in a fully disordered A2 structure. The particle's mean size was found to be ~23 nm with a dispersion of ± 10 nm. These NPs exhibit a soft ferromagnetic (FM) behavior with a very high $M_s$ and a high $T_c$. Interestingly, the low temperature (5K) magnetic moment per cell of the present disordered nanoscale system is around 6 $\mu_B$, which is in good agreement with the Slater-Pauling (SP) rule, indicating that the HMF is preserved in the A2 disordered CFG-NP system. It is widely known that the explanation of experimental results of HAs without first-principles density functional theory computations is quite difficult. Therefore, we also study the ground state (@T = 0 K) properties of the $Co_2FeGe$ system in its ideal $L2_1$ phase (space group # 225) via the first-principles calculation and compare them with our experimental results. Our experimental results are in close agreement with the theoretical predictions. The present study will illuminate some of the following key points:

(a) A new Heusler based nanoalloy will come into the mainstream of nanotechnology.

(b) It will increase our understanding of the unusual crystal structure of HAs at the nanoscale, exhibiting different physical properties compared with their bulk counterpart.

(c) Understanding of such Heusler based novel nanomaterials might open the way for diverse technological applications.

## 2. Methodology

### 2.1. Computational method

The ground-state properties of the $Co_2FeGe$ system are studied using the first-principles-based computational code Wien2K [52]. Previous studies established that the generalized gradient approximation (GGA) functional gives exact results, especially in systems having d and f orbitals, such as half-metallic Heusler alloys [13, 53]. For that reason, we use GGA functional of Perdew Burke Ernzerhof [54] in our calculations. The muffin-tin (MT) radii ($R_{MT}$) for Co, Fe, and Ge are 2.29, 2.29, and 2.23 a.u., respectively, ensuring that the MT spheres are nearly touching. The value of $R_{MT} \times K_{max}$ has been fixed at 7 throughout the calculations, where $K_{max}$ is the largest wave vector in the basis set. To achieve a good accuracy in results, large k-meshes (14 × 14 × 14, equivalent to 3000 k-points in the first Brillouin zone) are used. The spin-orbit coupling (SOC) effect has also been considered in the present study. For structural optimizations, the starting lattice constant 5.73 Å was taken from the X-ray diffraction (XRD) results. The ground state equilibrium parameters, viz., the equilibrium lattice constant ($a_0$), the total energy ($E_0$), and the bulk modulus ($B_0$) were determined by fitting the energy versus unit cell volume curve with the Birch-Murnaghan equation of state [55].

## 2.2. Experimental method

Coprecipitation and thermal deoxidization methods, as described in ref. [41], with some changes were used in the preparation of the $Co_2FeGe$ NPs. All the chemicals were purchased from Sigma-Aldrich, and were used as received. The precursors were Fe $(NO_3)_3 \cdot 9H_2O$ (99%), $CoCl_2 \cdot 6H_2O$ (99%), and $GeCl_4$ (99%). In a typical sample preparation procedure, these precursors were taken in an appropriate weight ratio (2.012 g, 2.371 g and 1.068 g respectively), and then were dispersed in 200 ml methanol ($CH_3OH$), followed by sonication for up to 30 minutes. Note that we did not use fumed silica as a template for controlling the nanoparticles growing process as described in ref. [41]. The obtained solution was further dried for the next 10 hours to collect the powder residue. The resultant powder was annealed in a tube furnace at 850 °C for 15 hours under $H_2$ environment. Finally, the product was placed inside glass vials to prevent it from oxidation and hydrolysis. The powder XRD with Cu Kα (λ=1.54 Å) radiation was performed to identify the phase of CFG-NPs. Microstructural properties were studied using a field-emission scanning electron microscope (FESEM) and a high-resolution transmission electron microscope (HRTEM). The elemental composition and purity of the sample were determined using energy-dispersive analysis of X-rays (EDAX). A physical property measurement system (PPMS) was employed to study the temperature dependent magnetic properties of the CFG-NPs. High-temperature (HT) magnetic properties were also studied using a Lakeshore vibrating sample magnetometer (VSM).

## 3. Results and discussion

### 3.1. Theoretical results

### 3.1.1. Structural properties

$X_2YZ$ type (X/Y: transition metals; Z: main group element) compounds such as $Co_2FeGe$ Heusler alloy (HA) can be crystallized in the conventional $L2_1$ phase of $Cu_2MnAl$-prototype [15], which comes under $F m\bar{3} m$ space group (# 225). This allows three Wyckoff sites: 8c (0.25, 0.25, 0.25) and (0.75, 0.75, 0.75); 4b (0.5, 0.5, 0.5) and 4a (0, 0, 0) [57], for the Co, Fe and Ge atoms, respectively, as shown in Fig. 1. To study the ground state properties of $Co_2FeGe$, volume optimizations are performed in its ferromagnetic (FM) state, and a change of total energy with respect to the cell volume (E-V) curve is shown in Fig. 2. The equilibrium lattice constant ($a_0$) and other ground state parameters as obtained after fitting the data with Birch Murnaghan equation of state are shown in Table 1. The optimized lattice constant ($a_0$) is found to be 5.748 Å, in agreement with the previously calculated values (see Table 1 for a comparison).

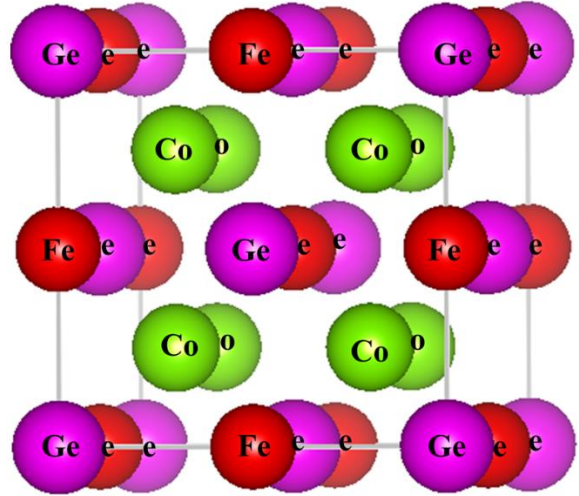

**Fig. 1.** $Co_2FeGe$ Heusler alloy crystal structure in ideal $L2_1$ or $Cu_2MnAl$-prototype phase, as generated using the XCrysden software [56].

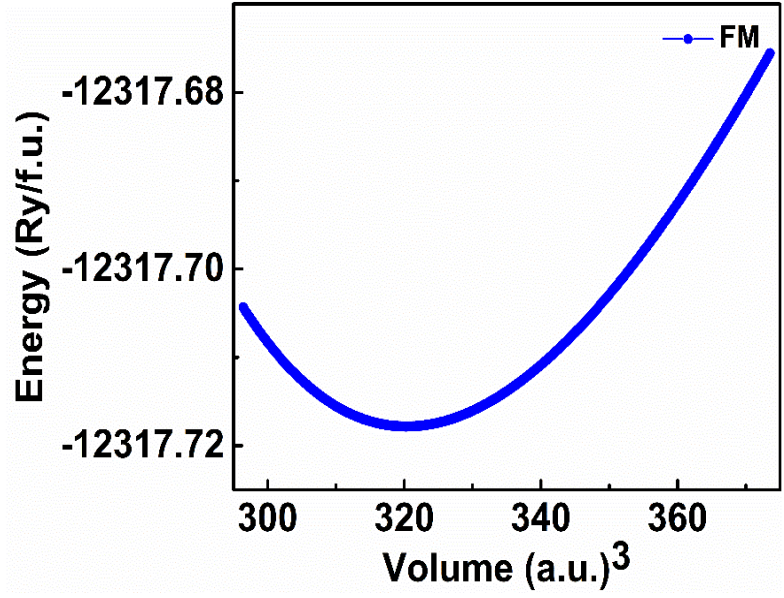

**Fig. 2.** Total energy versus unit cell volume curve for the ferromagnetic $Co_2FeGe$ alloy.

**Table 1**
The equilibrium lattice constant ($a_0$), bulk modulus (B), equilibrium volume ($V_0$), and equilibrium energy ($E_0$) per unit cell of $Co_2FeGe$.

| Parameters (@T=0K) | This work | Other work |
| --- | --- | --- |
| Lattice constant $a_o$ ( Å) | 5.748 | 5.758 [58]; 5.750 [59] |
| Bulk modulus B (GPa) | 190.556 | 162.677 [58]; |
| Derivative of Bulk modulus (B′) | 4.940 | - |
| Equilibrium volume ($V_0$) | 320.326 | - |
| Equilibrium energy ($E_0$) | -12317.718 | -12317.674 [58] |

### 3.1.2. Electronic structure and magnetic properties

The electronic structure and magnetic properties of $Co_2FeGe$ are studied at the optimized lattice constant 5.748 Å under a generalized gradient approximation (GGA) scheme. The total and atom-specific density of states (DOS) are presented in Fig. 3. Co and Fe $d$-states lead the DOSs near the $E_F$. The gap-like feature (at about 0.4 eV below the $E_F$) are observed in the spin-down band. The self-consistent field (SCF) calculation results are presented in Table 2. A total magnetic moment ($M_{Tot}$) of 5.64 $\mu_B$/f. u. is observed, which further enhances to ~ 5.72 $\mu_B$/f. u. upon consideration of the spin-orbit (SO) coupling effect during the SCF calculations. This enhanced value is closer to the SP value ($M_{SP}$) of 6 $\mu_B$. It has contributions primarily from the moments $M_{Co}$ and $M_{Fe}$ from the Co and Fe atoms, respectively, while the contributions from Ge ($M_{Ge}$) and interstitials ($M_{int.}$)

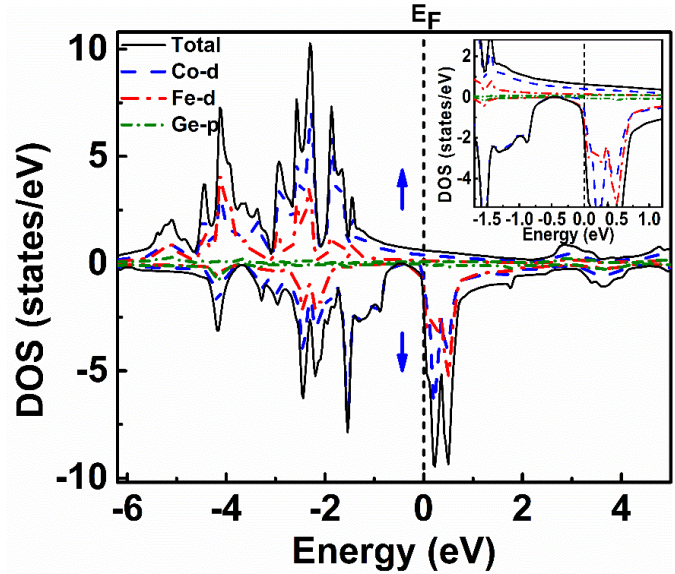

**Fig. 3.** The total and atom-specific DOSs for $Co_2FeGe$. The up and down arrows indicate the spin-up and spin-down bands, respectively. The Fermi energy is shown by a vertical dotted line. The inset shows the DOSs near the Fermi energy.

are negligible. The reason for the spin-down energy gap in $X_2YZ$ type systems is explained in our previous report [13]. The Curie temperature $T_c$, as calculated using the relation $T_c(K) = 23 + 181 \times M_{tot}$ ($\mu_B$/f.u.), as reported by Wurmehl *et al*. [1] is 1058 K.

**Table 2**

The calculated atoms specific, interstitial, and total magnetic moment per formula unit with and without the spin-orbit coupling effect.

| Spin polarized calculation results | This work | | Other work |
|---|---|---|---|
| | Without SO | With SO | |
| $M_{Co}(\mu_B)$ | 1.407 | 1.427 | 1.30 [59]; 1.421 [60] |
| $M_{Fe}(\mu_B)$ | 2.894 | 2.925 | 2.90 [59]; 2.916 [60] |
| $M_{Ge}(\mu_B)$ | 0.007 | 0.009 | 0.0 [59]; 0.010 [60] |
| $M_{int.}(\mu_B)$ | -0.073 | -0.072 | - |
| $M_{Tot}(\mu_B)$ | 5.643 | 5.716 | 5.61 [59]; 5.639 [60] |
| Slater Pauling value ($M_{sp}$) | | 6 ($\mu_B$/f.u.) [60] | |

## 3.2. Experimental results

### 3.2.1. Structural characteristics

The Co$_2$FeGe alloy, a member of full-HAs (of type X$_2$YZ) family, can be crystallized in L2$_1$ (well ordered), B2 (partially disordered) and in A2 (fully disordered) phases [17, 61, 62]. These phases are dependent upon the atomic occupations of the three Wyckoff sites (8c, 4b and 4a). The well-ordered L2$_1$ phase is obtained only when the cobalt (Co), iron (Fe), and germanium (Ge) atoms do not exchange their positions, i.e., they occupy their Wyckoff sites as shown in Fig. 1 and exhibit (111), (200) superlattice reflections in the XRD pattern [62]. In the case of partially disordered B2 phase, the (111) reflection is absent in XRD. While, for the fully disordered A2 phase wherein all the constituent elements are occupied randomly at the three sites, both the superlattice peaks (111) and (200) must be absent in the XRD pattern.

The experimental and simulated XRD patterns of the CFG-NPs are presented in Fig. 4. The prominent Bragg peaks (220), (400) and (422) observed at 44.67°, 64.97° and 82.34°, respectively, establish that the present sample is crystallized in a fully disordered phase of A2-type [16]. The lattice constant, as calculated using the most prominent (220) peak is 5.73 Å.

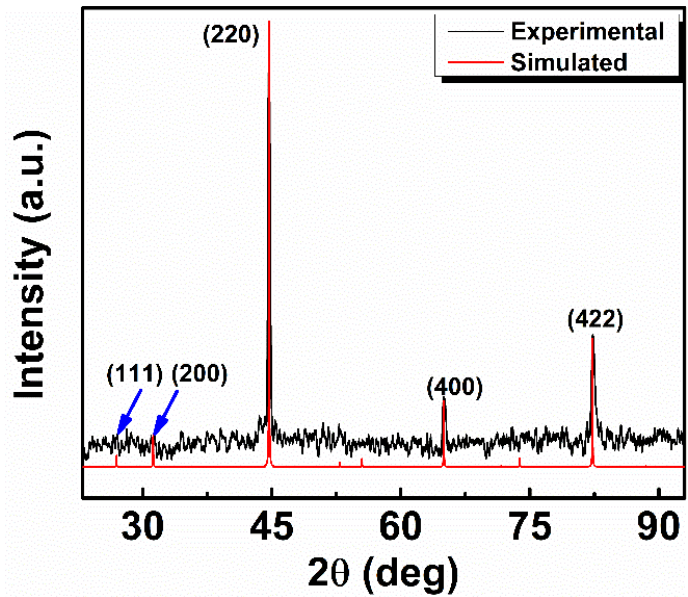

**Fig. 4.** Experimental and simulated XRD patterns of Co$_2$FeGe Heusler alloy nanoparticles annealed at 850 °C for 15 hours.

This closely matches with the bulk value [59, 63], and is also in a close agreement with our theoretically calculated equilibrium lattice constant 5.748 Å. The FESEM micrograph as shown in Fig. 5 (a) reveals that the nanoparticles are highly agglomerated. This suggests that the particles are highly magnetic. The EDAX spectrum, as shown in Fig. 5 (b), reveals that Co, Fe, and Ge are

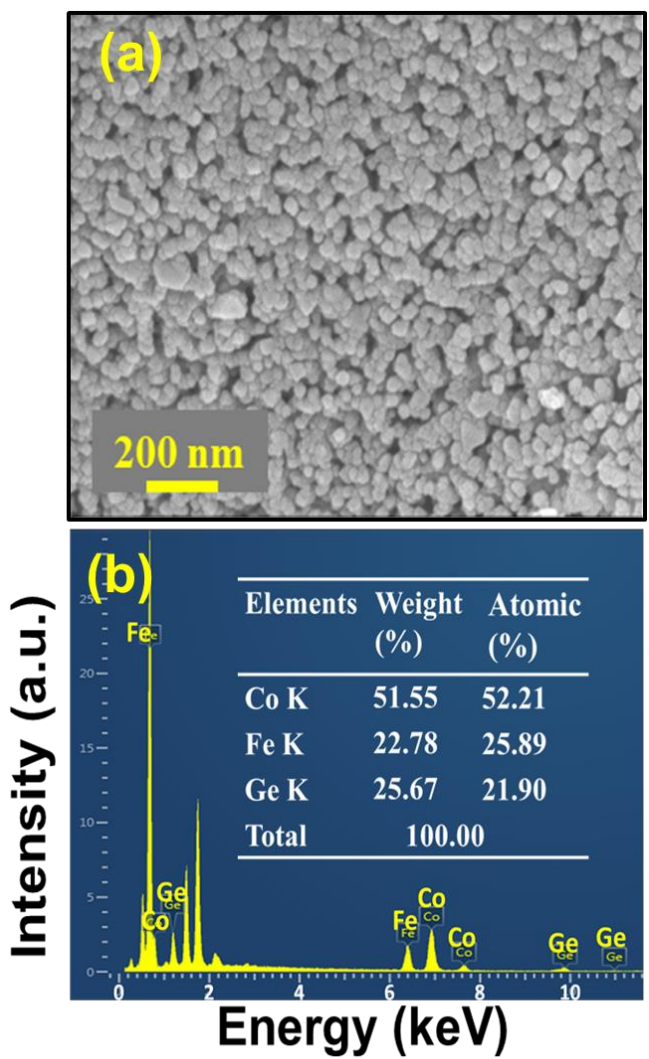

**Fig. 5** (a) FESEM image of Co$_2$FeGe-NPs annealed at 850 °C for 15 hours, and (b) EDAX spectrum. The inset shows the elemental composition of the nanoparticles.

the only elements in the NPs; the extra peaks are from the Si substrate and the gold coating. The atomic percentages as determined by EDAX and shown in the inset of Fig. 5 (b) are close to the composition Co$_2$FeGe, as desired. The HRTEM image shown in Fig. 6 (a) suggests that the NPs are nearly spherical in shape. The image has been analyzed for particle size distribution using the ImageJ software. The size-distribution histogram, along with the fitted Gaussian profile, is shown in Fig. 6 (b). According to the histogram, the particles are of sizes 23 ± 10 nm. The selected area electron diffraction image, as shown in Fig. 6 (c), features concentric rings along with dots. This

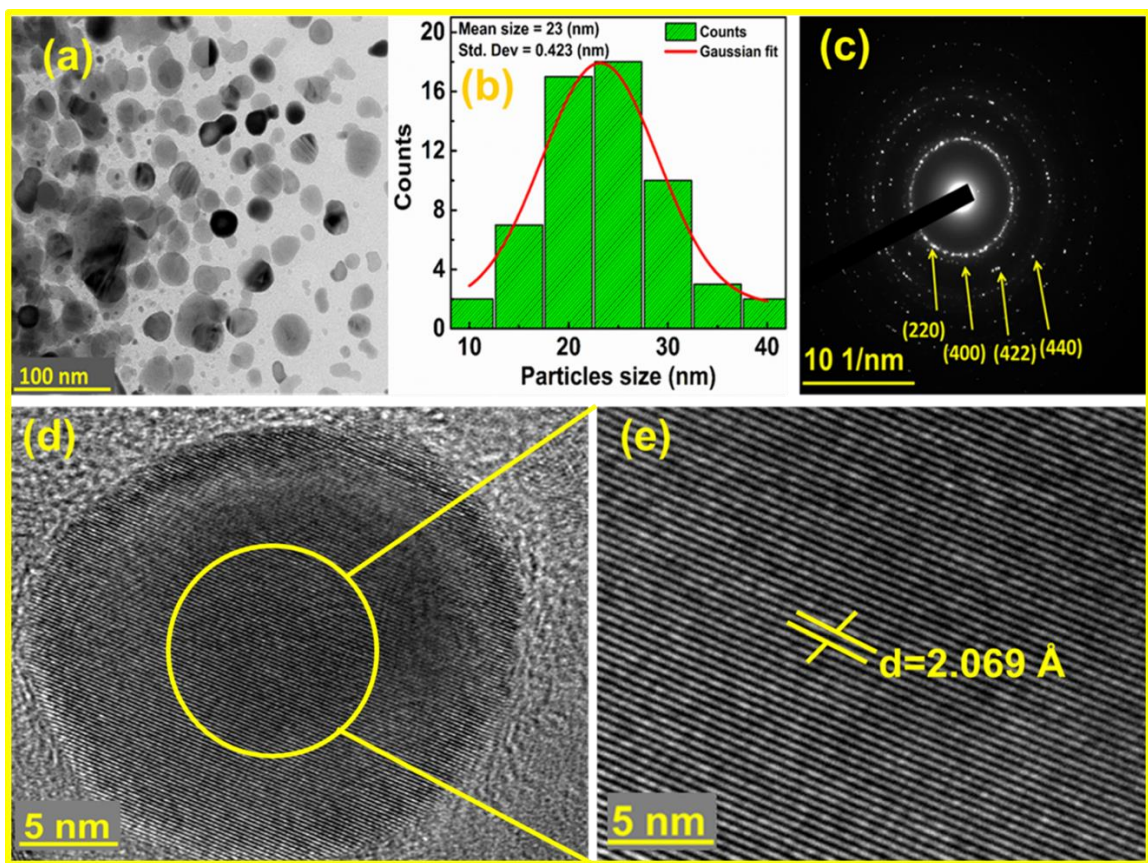

**Fig. 6** (a) TEM micrograph of the CFG-NPs. (b) A histogram for the size distribution. (c) SAED pattern with indexing of first four rings. (d, e) High-resolution images showing the lattice planes and crystallinity.

indicates that the NPs are crystalline in nature. The first four rings can be indexed with the (hkl) values (220), (400), (422) and (440), and thus are consistent with the XRD results. The image of a single particle of ~ 24 nm diameter with an even higher resolution is shown in Fig. 6 (d). The clearly visible lattice fringes confirm the high crystallinity of the NPs. An enlarged portion of the image is shown in Fig. 6 (e). The interplanar spacing of 2.069 Å, as shown, is equivalent to the (220) plane of the cubic Heusler phase of $Co_2FeGe$.

### 3.2.2. Magnetic properties

The field-dependent magnetization M (H) curves were taken at different temperatures from 5 to 300 K, as shown in Fig. 7. The low-temperature (5 K) saturation magnetization ($M_s$) is found to be 132.96 emu/g, which is equivalent to 5.9 $\mu_B$/f.u. This is quite close to both the computed value of 5.72 $\mu_B$/f.u. and value of 6.1 $\mu_B$/f.u. [64]. To become a perfect half-metal (100% spin-polarization), the total magnetic moment per unit cell should be an integer (e.g., 6 $\mu_B$), as predicted by the SP rule [65]. Interestingly, our experimental value (5.9 $\mu_B$/f.u.) closely matches with the SP rule, which further suggests that half-metallic ferromagnetism (HMF) is retained even at nanometric scale and in disordered phase. As is evident from the upper inset of Fig. 7, the finite values of the coercivity ($H_c$) and the remanence ($M_r$) indicate a soft ferromagnetic behavior of the Co$_2$FeGe nanoparticles. A comparison of the values of saturation magnetization ($M_s$), remanence ($M_r$), and coercivity ($H_c$) at different temperatures (5 - 300 K) are shown in Table 3. The temperature dependence of $M_s$, as presented in the lower inset of Fig. 7, reveals that the saturation magnetization persists even up to the room temperature, which is highly desirable for spintronic applications.

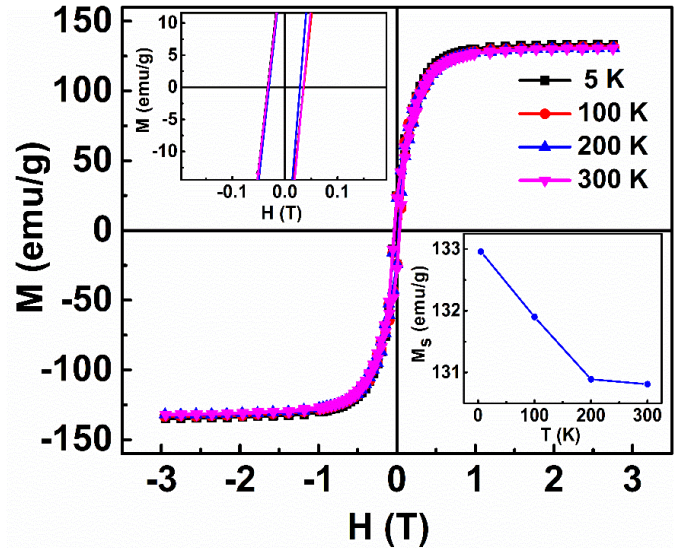

**Fig. 7.** Field dependence of magnetization curves of the $Co_2FeGe$ NPs taken at different temperatures. The upper inset displays the zoomed in view of the curve near zero-field, and the lower inset shows the temperature dependence of saturation magnetization.

**Table 3**

$M_s$, $M_r$, and $H_c$ at different temperatures (5 - 300 K), and Curie temperature ($T_c$) of Co$_2$FeGe NPs.

| Temperature (K) | Sat. magn. $M_s$ (emu/g) | Remanence $M_r$ (emu/g) | Coercivity $H_c$ (Oe) | Curie temperature (K) |
|---|---|---|---|---|
| 5 | 132.96 | 24.02 | 339.5 | 1060 [this work] |
| 100 | 131.90 | 23.77 | 332.5 | 1000 [66]; 981 [59]; |
| 200 | 130.89 | 23.05 | 294.5 | 1073 [64] |
| 300 | 130.81 | 22.49 | 336.5 | |
| **Bulk Sat. magn.** | | 6.1 $\mu_B$/f.u. [64] | | |
| **($M_s$) value** | | 5.61 $\mu_B$/f.u. (Theoretical) [59] | | |
| **Slater Pauling value** | | 6.0 $\mu_B$/f.u. [65] | | |

The temperature dependence of magnetization M (T) curve taken at 100 Oe under zero field-cooled (ZFC) condition is presented in Fig. 8 (a). The curve shows a ferromagnetic (FM) to paramagnetic (PM) phase transition. The $T_c$ is calculated from the first derivative of M (T), as shown in the inset of Fig. 8 (a), and is found to be 1060 K. This closely matches with the theoretically calculated value of 1058 K following the model provided by Wurmehl *et al.* [1]. The ZFC and field-cooled (FC) M (T) curves under the applied fields of 100 Oe and 20 kOe are presented in Fig. 8 (b). At 100 Oe field, a strong bifurcation between the ZFC and FC curves is observed. Such behavior is recognized as thermomagnetic irreversibility, and has previously been reported in various compounds [67-69]. It was reported that such behavior might be due to (i) the complex nature of the phase transition around and above the $T_c$, or (ii) due to non-collinear nature of the magnetic structure arising due to crystal filed effects, that could occur in highly anisotropic magnetic compounds, or (iii) due to domain wall pinning [70]. This irreversibility in magnetization is vanished under the application of the high 20 kOe field.

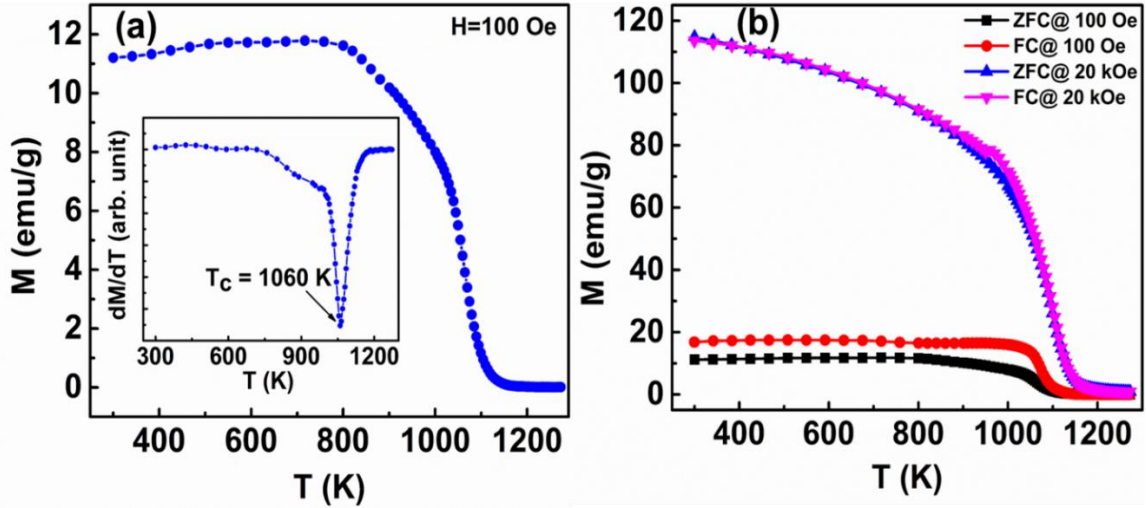

**Fig. 8.** (a) The temperature-dependent magnetization M (T) curve of $Co_2FeGe$ Heusler alloy nanoparticles at 100 Oe field. The inset displays the dM/dT versus T curve. (b) The temperature dependence of ZFC and FC magnetizations under 100 Oe and 20 kOe fields.

The temperature dependence of magnetization reveals that the ferromagnetic (FM) properties of the CFG-NP system are very stable. Further, it has a high saturation magnetization and a high $T_c$, indicating that it might be suitable for spin-based nanoscale devices.

## 4. Conclusions

Here, we report the synthesis and structural and magnetic properties of $Co_2FeGe$ nanoparticles with an average size of 23 nm. We have significantly facilitated a reported synthesis process, and have succeeded in reducing the particle size to this size with a low dispersion. The XRD pattern, and the HRTEM and SAED images have confirmed that the NPS are of the disordered A2 phase, and are crystalline in nature. These NPs exhibit a soft ferromagnetic behavior with a very high $M_s$, and a high $T_c$ of 1060 K. The low temperature (5 K) $M_s$ (5.9 $\mu_B$/f.u.) of the nanoparticles is in a good agreement with the predictions of the SP rule (an integer value), and indicates that the half-metallic ferromagnetism is preserved in particles even on the 20 nm length scale. The high $M_s$ and $T_c$ values suggest that the $Co_2FeGe$ nanoparticles are utilizable in spin-based nano-devices. Our experimental results are in close agreement with the results obtained from first-principles density functional theory computations.

**Acknowledgments**

A. Ahmad acknowledges University Grants Commission, New Delhi and Ministry of Education (MoE), India for providing the research fellowship. Amal Kumar Das acknowledges the financial support from Department of Science and Technology (DST), India (project no. EMR/2014/001026).

**Declaration of competing interest**

The authors declare that they have no known competing financial interests.

**References**

[1] S. Wurmehl, G.H. Fecher, H.C. Kandpal, V. Ksenofontov, C. Felser, H.J. Lin, J. Morais, Geometric, electronic, and magnetic structure of Co$_2$FeSi: Curie temperature and magnetic moment measurements and calculations, Phys. Rev. B 72 (2005) 184434.

[2] A.H. Lu, E.L. Salabas, F. Schüth, Magnetic nanoparticles: synthesis, protection, functionalization, and application, Angewand. Chem. Internat. Edition 46 (2007) 1222-1244.

[3] G. Ennas, A. Falqui, S. Marras, C. Sangregorio, G. Marongiu, Influence of metal content on size, dispersion, and magnetic properties of iron-cobalt alloy nanoparticles embedded in silica matrix, Chem. Mater. 16 (2004) 5659-5663.

[4] A. Corrias, M.F. Casula, A. Falqui, G. Paschina, Evolution of the structure and magnetic properties of FeCo nanoparticles in an alumina aerogel matrix, Chem. Mater. 16 (2004) 3130-3138.

[5] G. Ennas, A. Falqui, G. Paschina, G. Marongiu, Iron-Cobalt Alloy Nanoparticles Embedded in an Alumina Xerogel Matrix, Chem. Mater. 17 (2005) 6486-6491.

[6] G. Reiss, A. Hütten, Applications beyond data storage, Nat. Mater. 4 (2005) 725-726.

[7] S. Wei, Q. Wang, J. Zhu, L. Sun, H. Lin, Z. Guo, Multifunctional composite core–shell nanoparticles, Nanoscale 3 (2011) 4474-4502.

[8] G. Baldi, D. Bonacchi, C. Innocenti, G. Lorenzi, C. Sangregorio, Cobalt ferrite nanoparticles: The control of the particle size and surface state and their effects on magnetic properties, J. Magn. Magn. Mater. 311 (2007) 10-16.

[9]  B. Sohn, R. Cohen, G. Papaefthymiou, Magnetic properties of iron oxide nanoclusters within microdomains of block copolymers, J. Magn. Magn. Mater. 182 (1998) 216-224.

[10] T. Graf, C. Felser, S.S. Parkin, Simple rules for the understanding of Heusler compounds, Progr. Solid State Chem. 39 (2011) 1-50.

[11] M. Jourdan, J. Minár, J. Braun, A. Kronenberg, S. Chadov, B. Balke, A. Gloskovskii, M. Kolbe, H.J. Elmers, G. Schönhense, Direct observation of half-metallicity in the Heusler compound $Co_2MnSi$, Nat. Commun. 5 (2014) 1-5.

[12] C. Felser, L. Wollmann, S. Chadov, G.H. Fecher, S.S. Parkin, Basics and prospective of magnetic Heusler compounds, APL Materials 3 (2015) 041518.

[13] A. Ahmad, S.K. Srivastava, A.K. Das, Phase stability and the effect of lattice distortions on electronic properties and half-metallic ferromagnetism of $Co_2FeAl$ Heusler alloy: An ab initio study, J. Phys. Condens. Matter 32 (2020) 415606.

[14] A. Ahmad, A.K. Das, S.K. Srivastava, Competition of $L2_1$ and XA ordering in $Fe_2CoAl$ Heusler alloy: a first-principles study, Europ. Phys. J. B 93 (2020) 1-7.

[15] A. Ahmad, S. Srivastava, A. Das, Effect of $L2_1$ and XA ordering on phase stability, half-metallicity and magnetism of $Co_2FeAl$ Heusler alloy: GGA and GGA+ U approach, J. Magn. Magn. Mater. 491 (2019) 165635.

[16] A. Ahmad, S. Mitra, S.K. Srivastava, A.K. Das, Size-dependent structural and magnetic properties of disordered $Co_2FeAl$ Heusler alloy nanoparticles, J. Magn. Magn. Mater. 474 (2019) 599-604.

[17] B. Balke, S. Wurmehl, G.H. Fecher, C. Felser, J. Kübler, Rational design of new materials for spintronics: $Co_2FeZ$ (Z= Al, Ga, Si, Ge), Scie. Techno. Advanc. Mater. 9 (2008) 014102.

[18] R. Kainuma, Y. Imano, W. Ito, Y. Sutou, H. Morito, S. Okamoto, O. Kitakami, K. Oikawa, A. Fujita, T. Kanomata, Magnetic-field-induced shape recovery by reverse phase transformation, Nature 439 (2006) 957-960.

[19] R. Kainuma, K. Oikawa, W. Ito, Y. Sutou, T. Kanomata, K. Ishida, Metamagnetic shape memory effect in NiMn-based Heusler-type alloys, J. Mater. Chem. 18 (2008) 1837-1842.

[20] J. Barth, G.H. Fecher, B. Balke, S. Ouardi, T. Graf, C. Felser, A. Shkabko, A. Weidenkaff, P. Klaer, H.J. Elmers, Itinerant half-metallic ferromagnets $Co_2Ti\,Z$ (Z= Si, Ge, Sn): Ab initio calculations and measurement of the electronic structure and transport properties, Phys. Rev. B 81 (2010) 064404.

[21] S. Bhattacharya, A. Pope, R. Littleton IV, T.M. Tritt, V. Ponnambalam, Y. Xia, S. Poon, Effect of Sb doping on the thermoelectric properties of Ti-based half-Heusler compounds, $TiNiSn_{1-x}Sb_x$, Appl. Phys. Lett. 77 (2000) 2476-2478.

[22] S. Chadov, X. Qi, J. Kübler, G.H. Fecher, C. Felser, S.C. Zhang, Tunable multifunctional topological insulators in ternary Heusler compounds, Nat. Mater. 9 (2010) 541-545.

[23] X. L. Qi, R. Li, J. Zang, S.-C. Zhang, Inducing a magnetic monopole with topological surface states, Science 323 (2009) 1184-1187.

[24] H. Lin, L.A. Wray, Y. Xia, S. Xu, S. Jia, R.J. Cava, A. Bansil, M.Z. Hasan, Half-Heusler ternary compounds as new multifunctional experimental platforms for topological quantum phenomena, Nat. Mater. 9 (2010) 546-549.

[25] X. Zhang, H. Zhang, M. Qian, L. Geng, Enhanced magnetocaloric effect in Ni-Mn-Sn-Co alloys with two successive magnetostructural transformations, Sci. Rep. 8 (2018) 1-11.

[26] T. Krenke, E. Duman, M. Acet, E.F. Wassermann, X. Moya, L. Mañosa, A. Planes, Inverse magnetocaloric effect in ferromagnetic Ni–Mn–Sn alloys, Nat. Mater. 4 (2005) 450-454.

[27] T. Kojima, S. Kameoka, A.P. Tsai, The emergence of Heusler alloy catalysts, Scie. Technol. Advanc. Mater. 20 (2019) 445-455.

[28] T. Kojima, S. Kameoka, A.P. Tsai, Catalytic Properties of Heusler Alloys for Steam Reforming of Methanol, ACS Omega 4 (2019) 21666-21674.

[29] W. Wang, M. Przybylski, W. Kuch, L. Chelaru, J. Wang, Y. Lu, J. Barthel, H. Meyerheim, J. Kirschner, Magnetic properties and spin polarization of $Co_2MnSi$ Heusler alloy thin films epitaxially grown on GaAs (001), Phys. Rev. B 71 (2005) 144416.

[30] L. Singh, Z. Barber, A. Kohn, A. Petford-Long, Y. Miyoshi, Y. Bugoslavsky, L. Cohen, Interface effects in highly oriented films of the Heusler alloy $Co_2MnSi$ on GaAs (001), J. Appl. Phys. 99 (2006) 013904.

[31] T. Iwase, Y. Sakuraba, S. Bosu, K. Saito, S. Mitani, K. Takanashi, Large interface spin-asymmetry and magnetoresistance in fully epitaxial $Co_2MnSi/Ag/Co_2MnSi$ current-perpendicular-to-plane magnetoresistive devices, Appl. Phys. Expr. 2 (2009) 063003.

[32] J. Sato, M. Oogane, H. Naganuma, Y. Ando, Large magnetoresistance effect in epitaxial $Co_2Fe_{0.4}Mn_{0.6}Si/Ag/Co_2Fe_{0.4}Mn_{0.6}Si$ devices, Appl. Phys. Expr. 4 (2011) 113005.

[33] C. Wang, J. Meyer, N. Teichert, A. Auge, E. Rausch, B. Balke, A. Hütten, G.H. Fecher, C. Felser, Heusler nanoparticles for spintronics and ferromagnetic shape memory alloys, J. Vacuum Sci. Techn. B 32 (2014) 020802.

[34] V.V. Khovaylo, V.V. Rodionova, S.N. Shevyrtalov, V. Novosad, Magnetocaloric effect in "reduced" dimensions: Thin films, ribbons, and microwires of Heusler alloys and related compounds, Physica Status Solidi b 251 (2014) 2104-2113.

[35] Y. Zhang, F. Qin, D. Estevez, V. Franco, H. Peng, Structure, magnetic and magnetocaloric properties of $Ni_2MnGa$ Heusler alloy nanowires, J. Magn. Magn. Mater. 513 (2020) 167100.

[36] C.I. Tsai, C.Y. Wang, J. Tang, M.H. Hung, K.L. Wang, L.J. Chen, Electrical properties and magnetic response of cobalt germanosilicide nanowires, ACS Nano 5 (2011) 9552-9558.

[37] P. Simon, D. Wolf, C. Wang, A.A. Levin, A. Lubk, S. Sturm, H. Lichte, G.H. Fecher, C. Felser, Synthesis and three-dimensional magnetic field mapping of $Co_2FeGa$ Heusler nanowires at 5 nm resolution, Nano Lett. 16 (2016) 114-120.

[38] S. Khan, N. Ahmad, N. Ahmed, A. Safeer, J. Iqbal, X. Han, Structural, magnetic and transport properties of Fe-based full Heusler alloy $Fe_2CoSn$ nanowires prepared by template-based electrodeposition, J. Magn. Magn. Mater. 465 (2018) 462-470.

[39] N. Ahmad, N. Ahmed, X. Han, Analysis of electronic, magnetic and half-metallic properties of $L2_1$-type ($Co_2Mn_{0.5}Fe_{0.5}Sn$) Heusler alloy nanowires synthesized by AC-electrodeposition in AAO templates, J. Magn. Magn. Mater. 460 (2018) 120-127.

[40] L. Galdun, V. Vega, Z. Vargová, E.D. Barriga-Castro, C. Luna, R. Varga, V.M. Prida, Intermetallic $Co_2FeIn$ heusler alloy nanowires for spintronics applications, ACS Appl. Nano Mater. 1 (2018) 7066-7074.

[41] L. Basit, C. Wang, C.A. Jenkins, B. Balke, V. Ksenofontov, G.H. Fecher, C. Felser, E. Mugnaioli, U. Kolb, S.A. Nepijko, Heusler compounds as ternary intermetallic nanoparticles: $Co_2FeGa$, J. Phys. D Appl. Phys. 42 (2009) 084018.

[42] C. Wang, A.A. Levin, J. Karel, S. Fabbrici, J. Qian, C.E. ViolBarbosa, S. Ouardi, F. Albertini, W. Schnelle, J. Rohlicek, Size-dependent structural and magnetic properties of chemically synthesized Co-Ni-Ga nanoparticles, Nano Research 10 (2017) 3421-3433.

[43] F. Yang, J. Min, Z. Kang, S. Tu, H. Chen, D. Liu, W. Li, X. Chen, C. Yang, The influence of pH value and composition on the microstructure, magnetic properties of Co-Fe-Al Heusler nanoparticles, Chem. Phys. Lett. 670 (2017) 1-4.

[44] P. Nehla, C. Ulrich, R.S. Dhaka, Investigation of the structural, electronic, transport and magnetic properties of Co$_2$FeGa Heusler alloy nanoparticles, J. Alloys. Compd. 776 (2019) 379-386.

[45] A. Ahmad, S. Mitra, S.K. Srivastava, A.K. Das, Structural, magnetic, and magnetocaloric properties of intermetallic Fe$_2$CoAl Heusler nanoalloy, arXiv:2102.11195 (2021).

[46] A. Ahmad, S. Mitra, S.K. Srivastava, A.K. Das, Giant magnetocaloric effect in Co$_2$FeAl Heusler nanoalloy, arXiv:1909.10201 (2019).

[47] W. Duan, L. Yang, Y. Li, J. Guo, M. Song, The influence of solvents on the microstructure and magnetic properties of Co$_2$FeAl Heusler alloy nanoparticles, Mater. Chem. Phys. 256 (2020) 123724.

[48] G. Saravanan, V. Asvini, R. Kalaiezhily, K. Ravichandran, Effect on Annealing Temperature (Ta) of Ternary Full Fe$_2$CrSi Heusler Alloy Nanoparticles for Spin-Based Device Applications, J. Supercond. Nov. Magn. 33 (2020) 3957-3962.

[49] Y. Kobayashi, S. Tada, R. Kikuchi, Porous intermetallic Ni$_2$XAl (X= Ti or Zr) nanoparticles prepared from oxide precursors, Nanoscale Adv. 3 (2021) 1901-1905.

[50] G. Cavazzini, F. Cugini, D. Delmonte, G. Trevisi, L. Nasi, S. Ener, D. Koch, L. Righi, M. Solzi, O. Gutfleisch, Multifunctional Ni-Mn-Ga and Ni-Mn-Cu-Ga Heusler particles towards the nanoscale by ball-milling technique, J. Alloys Compd. 872 (2021) 159747.

[51] V. Asvini, G. Saravanan, R. Kalaiezhily, V. Ganesan, K. Ravichandran, Soft Ferromagnetic Properties of Half-Metallic Mn$_2$CoAl Heusler Alloy Nanoparticles for Spintronics Applications, J. Supercond. Nov. Magn. 33 (2020) 2759-2766.

[52] P. Blaha, K. Schwarz, G.K. Madsen, D. Kvasnicka, J. Luitz, R. Laskowsji, F. Tran, L. Marks, An Augmented Plane Wave+ Local Orbitals Program for Calculating Crystal Properties, Techn, Universitat Wien, Austria (2001).

[53] B. Fadila, M. Ameri, D. Bensaid, M. Noureddine, I. Ameri, S. Mesbah, Y. Al-Douri, Structural, magnetic, electronic and mechanical properties of full-Heusler alloys $Co_2YAl$ (Y= Fe, Ti): first principles calculations with different exchange-correlation potentials, J. Magn. Magn. Mater. 448 (2018) 208-220.

[54] J.P. Perdew, K. Burke, M. Ernzerhof, Generalized gradient approximation made simple, Phys. Rev. Lett. 77 (1996) 3865.

[55] F. Birch, Finite elastic strain of cubic crystals, Phys. Rev. 71 (1947) 809.

[56] A. Kokalj, Computer graphics and graphical user interfaces as tools in simulations of matter at the atomic scale, Comput. Mater. Sci. 28 (2003) 155-168.

[57] M. Zagrebin, V. Sokolovskiy, V. Buchelnikov, Electronic and magnetic properties of the $Co_2$-based Heusler compounds under pressure: first-principles and Monte Carlo studies, J. Phys. D Appl. Phys. 49 (2016) 355004.

[58] D. Rai, A. Shankar, M. Ghimire, R. Thapa, A comparative study of a Heusler alloy $Co_2FeGe$ using LSDA and LSDA+ U, Physica B Cond. Matter 407 (2012) 3689-3693.

[59] K.R. Kumar, K.K. Bharathi, J.A. Chelvane, S. Venkatesh, G. Markandeyulu, N. Harishkumar, First-principles calculation and experimental investigations on full-Heusler alloy $Co_2FeGe$, IEEE Trans. Magnetics 45 (2009) 3997-3999.

[60] N. Uvarov, Y. Kudryavtsev, A.F. Kravets, A.Y. Vovk, R. Borges, M. Godinho, V. Korenivski, Electronic structure, optical and magnetic properties of $Co_2FeGe$ Heusler alloy films, J. Appl. Phys. 112 (2012) 063909.

[61] S. Wurmehl, G.H. Fecher, K. Kroth, F. Kronast, H.A. Dürr, Y. Takeda, Y. Saitoh, K. Kobayashi, H.J. Lin, G. Schönhense, Electronic structure and spectroscopy of the quaternary Heusler alloy $Co_2Cr_{1-x}Fe_xAl$, J. Phys. D Appl. Phys. 39 (2006) 803.

[62] J. Du, Y. Zuo, Z. Wang, J. Ma, L. Xi, Properties of $Co_2FeAl$ Heusler alloy nano-particles synthesized by coprecipitation and thermal deoxidization method, J. Mater. Sci. Techn. 29 (2013) 245-248.

[63] B. Balke, S. Wurmehl, G.H. Fecher, C. Felser, M.C. Alves, F. Bernardi, J. Morais, Structural characterization of the $Co_2Fe$ Z (Z= Al, Si, Ga, and Ge) Heusler compounds by x-ray diffraction and extended x-ray absorption fine structure spectroscopy, Appl. Phys. Lett. 90 (2007) 172501.

[64] S. Mitra, A. Ahmad, S. Chakrabarti, S. Biswas, A.K. Das, Investigation on structural, electronic and magnetic properties of $Co_2FeGe$ Heusler alloy: experiment and theory, arXiv:2010.09590 (2020).

[65] G.H. Fecher, H.C. Kandpal, S. Wurmehl, C. Felser, G. Schönhense, Slater-Pauling rule and Curie temperature of $Co_2$-based Heusler compounds, J. Appl. Phys. 99 (2006) 08J106.

[66] B.C.S. Varaprasad, A. Srinivasan, Y. Takahashi, M. Hayashi, A. Rajanikanth, K. Hono, Spin polarization and Gilbert damping of $Co_2Fe(Ga_xGe_{1-x})$ Heusler alloys, Acta Materialia 60 (2012) 6257-6265.

[67] A. Murtaza, W. Zuo, M. Yaseen, A. Ghani, A. Saeed, C. Hao, J. Mi, Y. Li, T. Chang, L. Wang, Magnetocaloric effect in the vicinity of the magnetic phase transition in $NdCo_{2-x}Fe_x$ compounds, Phys. Rev. B 101 (2020) 214427.

[68] J. Panda, S. Saha, T. Nath, Critical behavior and magnetocaloric effect in $Co_{50-x}Ni_xCr_{25}Al_{25}$ (x= 0 and 5) full Heusler alloy system, J. Alloys Compd. 644 (2015) 930-938.

[69] T. Chabri, A. Venimadhav, T. Nath, Magnetic and lattice entropy change across martensite transition of Ni-Mn-Sn melt spun ribbons: Key factors in magnetic refrigeration, J. Magn. Magn. Mater. 466 (2018) 385-392.

[70] M. Foldeaki, A. Giguere, R. Chahine, T. Bose, Development of intermetallic compounds for use as magnetic refrigerators or regenerators, Advances Cryogenic Engineering (Springer) 3 (1998) 1533-1540.